\documentclass[aps, prd, amsmath, floats,floatfix, twocolumn, superscriptaddress, nofootinbib, showpacs]{revtex4}
\usepackage{amsmath} 
\usepackage{longtable}
\usepackage{verbatim} 
\usepackage{mathrsfs}
\usepackage{amsfonts}
\usepackage{dsfont}
\usepackage{color}
\usepackage{graphicx}

\newcommand{\be}{\begin{equation}}
\newcommand{\ee}{\end{equation}}
\newcommand{\bea}{\begin{eqnarray}}
\newcommand{\eea}{\end{eqnarray}}

\begin{document}

\title{Neutrino Spin Flavor Precession and Leptogenesis}
\author{Juan Barranco}
\affiliation{Division de Ciencias e Ingenier\'ias,  Universidad de Guanajuato, Campus Leon,
 C.P. 37150, Le\'on, Guanajuato, M\'exico.}

\author{Roberto Cota}
\affiliation{Division de Ciencias e Ingenier\'ias,  Universidad de Guanajuato, Campus Leon,
 C.P. 37150, Le\'on, Guanajuato, M\'exico.}

\author{David Delepine}
\affiliation{Division de Ciencias e Ingenier\'ias,  Universidad de Guanajuato, Campus Leon,
 C.P. 37150, Le\'on, Guanajuato, M\'exico.}

\author{Shaaban Khalil}
\affiliation{{\fontsize{10}{10}\selectfont{Centre for Theoretical Physics, Zewail City of Science and Technology, Sheikh Zayed, 12588, Giza, Egypt.}}}
\affiliation{{\fontsize{10}{10}\selectfont{Department of Mathematics, Faculty of Science, Ain Shams University, Cairo, Egypt.}}}

\date{\today}

\begin{abstract}
We argue that $\Delta L=2$ neutrino spin flavor precession,
induced by the primordial magnetic fields, could have a
significant impact on the leptogenesis process that accounts for
the baryon asymmetry of the universe. Although the extra galactic
magnetic fields is extremely weak at present time (about $10^{-9}$
Gauss), the primordial magnetic filed at the electroweak scale
could be quite strong (of order $10^{17}$ Gauss). Therefore, at
this scale, the effects of the spin flavor precession are not
negligible. We show that the lepton asymmetry may be reduced by
$50 \%$ due to the spin flavor precession. In addition, the
leptogenesis will have different feature from the standard scenario
of leptogenesis, where the lepton asymmetry continues to oscillate
even after the electroweak phase transition.

\end{abstract}
\maketitle

\section{Introduction}

Observations indicate with high level of accuracy that the present
universe contains no significant amount of baryonic antimatter
\cite{Steigman:1976ev,Steigman:2008ap}. Thus, the baryonic matter
we are made off is the remanent of a small matter-antimatter
asymmetry originated at the early universe. This asymmetry can not
be explained within both the Standard Model of Particle Physics
(SM) and the Standard Model of Cosmology, usually called Lambda
Cold Dark Matter ($\Lambda$CDM). Fortunately, an elegant
explanation of the observed baryon asymmetry is offered by
neutrino physics. This mechanism requires right-handed Majorana
neutrinos that decay out-of-equilibrium. This decay process,
combined with non-perturbative anomalous electro-weak processes,
can generate the baryon number in the universe
\cite{Fukugita:1986hr,Luty:1992un,Roulet:1997xa,Buchmuller:1997yu,Buchmuller:2004nz}.
In this 'leptogenesis' generation of the baryon asymmetry, it is
expected that the lepton asymmetry to be of the same order of
magnitude that of the baryon asymmetry, due to sphaleron effects
that are relevant for temperature from $10^{12}~\rm{GeV}$ to
$100~\rm{GeV}$ \cite{Kuzmin:1985mm,Matveev:1988pj}. The
measurement of the Baryon Asymmetry of the Universe (BAU) through
the anistropies of the cosmic microwave background radiation (CMB)
together with other cosmological observations at a very high level
of precision have strongly constrained BAU, that is parameterized
by the ratio of baryon number to photon number:
$\eta_B=N_B/N_\gamma$. Recent
analysis \cite{Steigman:2010zz} implies that %
\begin{equation}
\eta_B=(5.8\pm0.27) \times 10^{-10},
\end{equation}
which show that the measurement of baryon asymmetry is achieved
with an error less than 5\%.

In addition, the neutrino physics is also reaching high precision
measurements. Recently the last lepton mixing matrix angle
$\theta_{13}$ has been measured \cite{Ahn:2012nd}. These
progresses should permit to test leptogenesis models. So it is
very important to have a reliable way to describe the production
of the lepton asymmetry taking into account all $\Delta  L \neq 0$
processes. In fact, the lepton asymmetry is not precisely measured
as the baron asymmetry. Recently, it has been trying to constraint
the lepton asymmetry from WMAP and nucleosynthesis
\cite{castorina}. The following limits on
$\eta_L=(N_{\nu_L}-N_{\bar \nu_L})/N_{\gamma}$ have been obtained:
\begin{equation}
-0.071< \eta_L <0.054.
\end{equation}
It is clear that these limits are far from the accurate precision
of the baryon asymmetry. Nevertheless, new WMAP measurements and
better knowledge on neutrino mixing matrices would permit to
improve these results.

The fact that  non diagonal neutrino magnetic moment $\mu_\nu$
could induce a neutrino-antineutrino transition due to a helicity
flip produced by the interaction of $\mu_\nu$ with an external
magnetic field is known since a long time. It has been called
Spin-Flavor precession (SFP) effect. This effect was originally
used to explain the solar neutrinos deficit
\cite{Cisneros:1970nq,Okun:1986hi,Okun:1986na,Miranda:2000bi}.
However, after the confirmation of mixing mass explanation by
KamLAND \cite{Barranco:2002te}, the SFP is used as a mechanism to
constraint $\mu_\nu$ \cite{Miranda:2003yh}. In this letter, we
consider the implications of the neutrino SFP on $\Delta L=2$
processes and leptogenesis. To our knowledge, this is the first
time that this neutrino-antineutrino transition is analyzed in the
context of $\Delta L=2$ process that might affect the leptonic
asymmetry produced in early universe. The effect of a primordial
magnetic field on baryogenesis have already been studied
\cite{Giovannini:1997eg,Semikoz:2007ti,Semikoz:2009ye,Giovannini:1997eg}
but it has been done using the standard model anomaly terms which
violates $B+L$ quantum numbers and not through SFP process.

The letter is organized as follows. In section II  we briefly
review the neutrino spin flavor precession, induced by the
primordial magnetic fields. In section III the time dependent
magnetic fields at early universe is discussed. Section IV is
devoted for a possible lepton asymmetry generated by the SFP
process. In section V the associated leptogenesis induced by SFP
is studied. Finally our conclusions and remarks are given in
section VI.

\section{Neutrino Spin Flavor Precession}

The assumption that neutrino magnetic moment could be an explanation to the deficiency of the solar neutrino flux through Spin Precession effect were exposed by Cisneros more than 40 years ago \cite{Cisneros:1970nq}
and generalized later to the case of Majorana neutrinos
\cite{Schechter:1981hw}. It is well known that left-handed fermion with magnetic moment could be affected by the Spin Precession effect (SP) which induces in presence of magnetic field a transition from left to right handed fermions or inversely\cite{Akhmedov:1997yv}.
For the Majorana neutrinos the diagonal components of magnetic
moments vanish and the off-diagonal components are related by $-
\mu_{e \mu} = \mu_{\mu e} \equiv \mu_\nu $ leading to processes
violating flavors and lepton number. In order to find the
probability of the $\nu_{eL}\to{\nu}_{\mu L}^c$ transition, we
need to study the evolution of the chiral components of two
flavors of neutrinos, which is described by a Schr\"odinger type
equation \cite{Lim:1987tk}. In general, in a medium with arbitrary
matter density and magnetic field profiles, no analytical
closed-form expression for the transition probability can be
obtained. In this case one has to solve Schr\"odinger equation
numerically, which is quite straightforward. Since we are
interested at early epochs of the universe ($T \sim
10^{11}~\rm{GeV}$), we  ignore the neutrino masses and the
electron-neutron energy densities. Hence, the solution to the
Schr\"odinger type equation \cite{Lim:1987tk} with an arbitrary
magnetic field is given by \be P(\nu_{eL}\to {\nu}_{\mu
L}^c;t)=\sin^2 \left(\int_{t_0}^t\mu_\nu B_\bot(t')\,dt'\right)\,.
\label{P1} \ee where $B_\bot(t)$ is the transverse magnetic field
strength, $t$ being the time appearing in the Schr\"odinger
equation. This formula is valid for an arbitrary magnetic field
profile $B_\bot(t)$. It is interesting to note that Eq. (\ref{P1})
is valid after the electroweak breaking scale in a first
approximation due to the fact that the energy of the neutrino $E$
is given by the temperature $T$ which is much bigger than $\Delta
m^2$, so the terms proportional to $\Delta m^2/E$ can still be
neglected. It is important to notice that the SFP will not stop at
Electroweak breaking scale but will continue up to our days.
This can be understood from Eq. (\ref{P1}), where the probability
depends on the magnitude of the magnetic moment and time scale. It
is clear that at every times {\it i.e.}, $t$ is very small, the
magnetic field $B$ must be extremely large in order to imply a
large SFP effect. While at late time {\it i.e.}, $t$ is very
large, a reasonable SFP effect can be obtained with smaller values
of $B$. As an example let us consider a constant magnetic field of
order $10^4~G$, it is clear that the SFP effect is irrelevant
between the scale of right-handed neutrino decays ($t_{M_1\simeq
10^{11}GeV }\simeq 10^{-30}s$) and electroweak symmetry breaking
($t_{EPT} \simeq 10^{-11} s$). However, its effect becomes
important at time around the big bang nucleosynthesis (BBN).

It is remarkable that the spin flavor precession probability
violates the lepton number by two units ($\Delta L=2$). Therefore,
we can conjecture that such term may affect the Leptogenesis
scenario.

\section{Time-dependent magnetic fields at Early Universe}

The main constraints on SFP processes are coming from limits on
primordial magnetic field at photon decoupling time obtained
through observing microwave background radiation
\cite{Ichiki:2011ah} which puts a limit on present time magnetic
field to be smaller than $3\times 10^{-9}~G$ \cite{Grasso:2000wj}.
This limit should be translated into the primordial time assuming
that the magnetic field evolution is given by
\begin{equation}
B(t)\simeq B(t_i)\left(\frac{a(t_i)}{a(t)}\right)^2,
\label{123}
\end{equation}
 where $a(t)$ is the scale factor, assuming Friedman Robertson Walker dynamics for the Universe.
 Usually, the relation between the magnetic fields at different cosmological time is not so simple but for simplicity,
 we shall assume  this scaling factor (for detailed discussion see ref. \cite{Kandus:2010nw}).
 This means that  the CMB limit on present value of the primordial magnetic field could  be roughly translated into a limit
 of order $10^{9} G$ for the primordial magnetic field at BBN time \cite{Matese:1969zz,Kandus:2010nw},
 which correspond to a time of around $100~s$ after Big Bang
 \footnote{ In Ref. \cite{Caprini:2001nb}, it has been claimed that the limits on gravitational waves could
 bound much stronger than the BBN bounds.
 But in more recent studies \cite{Wang:2008vp}, it has been shown that  the limits on cosmological magnetic fields set
 by the latest LIGO S5 data lie close to those obtained by BBN and the CMB.
 For a review on Primordial Magnetic fields see ref. \cite{Kandus:2010nw}}.
 Thus, it is crucial to translate this bound on the primordial magnetic fields at electroweak symmetry breaking scale
 and up to the scale of right-handed majorana neutrino decoupling ($M_1 $ around $10^{11}GeV$), where the leptogenesis process takes place.
 At these times (which correspond to radiation domination era) , the scale factor is given by%
 \begin{equation}
 a(t) \propto t^{1/2}. %
 \end{equation}
 Thus, the bound on magnetic field at electroweak scale is of order $10^{17}~G$ and at
 $M_1$ scale is around $10^{27} G$.
 In this respect, we assume that our time-dependent magnetic field between the time associated to the scale of
 the heavy right-handed Majorana Neutrinos typically given by $M_1 \simeq 10^{11} GeV$
 and the time ($t_{EPT}$), which corresponds to the time when the Electoweak Phase Transition  (EPT) occurs, is given by:
 \begin{equation}
 B(t) \simeq B(t_{EPT}) \frac{t_{EPT}}{t}
 \label{Bt}
 \end{equation}
 where $B(t_{EPT}) \simeq 10^{17} G$.

\section{Spin Flavor Precession and Lepton asymmetry}

We now study the effect of spin-precession process on light,
mainly left-handed, majorana neutrino assuming the existence of a
time-dependent primordial magnetic fields given in Eq.(\ref{Bt}).
In order to include  in the Boltzman equation the terms
corresponding to the spin-precession effects, it is important to
recall in two flavor case that the variation $\Delta
N_{\nu_{1,2}}$  in  $\nu_{1,2}$ number density due to SFP is given
by
\begin{eqnarray}
\Delta N_{\nu_1} &=& P(\nu_2^c \rightarrow \nu_1)N_{\nu_2^c}-P(\nu_1 \rightarrow \nu_2^c)N_{\nu_1} \label{1}\\
\Delta N_{\nu_2} &=& P(\nu_1 \to \nu_2^c) N_{\nu_1} - P(\nu_2^c
\to \nu_1) N_{\nu_2^c}
\end{eqnarray}
where the first term in Eq. (\ref{1}) account for the number of
$\nu_2^c$'s which have been changed into $\nu_1$ and the second
term is equal to the number of $\nu_1$'s which have been changed
into $\nu_2^c$. Similar equations can be built for $\Delta
N_{\nu_{1,2}^c}$. Defining the lepton number density as %
$$ N_L \equiv N_{\nu_1} + N_{\nu_2}-N_{\nu_1^c} -N_{\nu_2^c}~,$$%
and assuming $CP$ is conserved ({\it i.e.}, $P(\nu_1 \to \nu_2^c)=P(\nu_1^c \to \nu_2)$), one gets %
\bea %
\Delta N_L = - 2 P N_L
\label{nL2} %
\eea%
where $P$ is the probability of SFP given by Eq. (\ref{P1}). This
approach can be easily extend to $n_f$ flavors and we obtain \bea
\frac{d N_L}{d t} &=& - 2 ( n_f - 1) \frac{d}{d t} \left( P  N_L
\right) . %
\label{NLFF} %
\eea%
This equation represents a new contribution for the lepton
asymmetry that will affect the leptogenesis scenario. Thus, the
lepton number density $N(t)$ is given by%
\be \label{case1} %
N_L(t)=\frac {N_L^0}{1 + 2 (n_f-1) P(\bar{\nu} \to \nu)}, %
\ee %
where $N_L^0$ is the initial lepton number density. It is
clear that for probability $P(\bar{\nu} \to \nu) \simeq {\cal
O}(1)$, the lepton asymmetry can be reducing respect to its
initial value by a factor $1/5$.

  It is worth mentioning that for magnetic fields below $10^{14}~G$
and due to limit on the neutrino magnetic moment
\cite{Miranda:2003yh} to be $\mu < 10^{-12} \mu_B$, where $\mu_B$
is the Bohr magneton, one can easily check from Eq.(\ref{P1}) that
the SFP process is irrelevant between the scale of right-handed
neutrino decays ($t_{M_1 }$) and electroweak symmetry breaking
time ($t_{EPT}$). So for primordial
magnetic field smaller than $10^{14}G$, SFP process will not
affect directly the usual leptogenesis scenario. However as it will be shown
explicitly below, it will continue to affect the
lepton asymmetry even after the electroweak symmetry breaking, transforming it as a time-oscillating function. This is important as in usual leptogenesis scenario, the lepton and baryon asymmetry of the Universe are related through a simple relation which only depends on matter contents of the model. Even with a relatively weak primordial magnetic field (below $10^{14}~G$), this relation between $\eta_L$ and $\eta_B$ is lost.

\section{Leptogenesis and Spin Flavor Precession}
\begin{figure}
  \centering
  \includegraphics[scale=0.30]{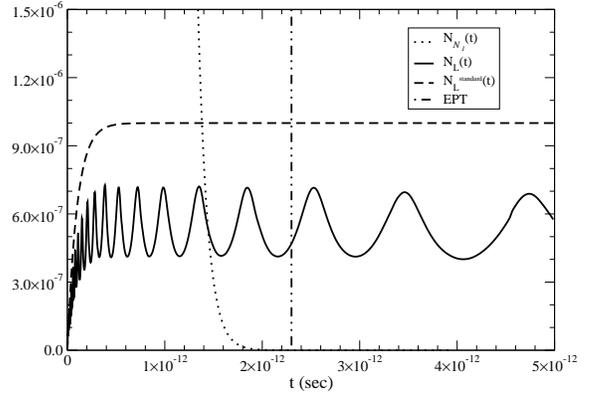}
  \caption{ The continue oscillating line corresponds to $\eta_L$ including SFP effects.
  The dash line is $\eta_L$ as expected from standard leptogenesis scenario.The vertical dotted-dash line
  represents an approximate value for  Electoweak Phase Transition Time ($t_{EPT}$). For $t>t_{EPT}$, the $\eta_L$
  is not anymore converted into $\eta_B$  through $B+L$ violating sphalerons. The dotted line show the evolution of the
  heavy RH majorana neutrino density $N_1$}
\label{fig2}
\end{figure}

Here, we assume a strong primordial time-dependent magnetic field,
given by Eq.(\ref{Bt}), before electroweak phase transition and
compatible with present limits on cosmological magnetic fields. In
order to get the SFP effects on Leptogenesis standard
scenario\footnote{ for a detail description of Leptogenesis
standard scenario see for instance ref. \cite{Buchmuller:2004nz}},
we solve  the Boltzman equation of the heavy  right-handed (RH)
majorana neutrino, $N_1$, that decays violating $CP$ and producing
a lepton asymmetry through usual leptogenesis scenario. This
lepton asymmetry is transformed into the Baryon Asymmetry of the
Universe through anomalous $B+L$ violating sphaleron processes
which are in equilibrium between $ 10^{12} GeV> T> 100~ GeV$. For
simplicity and in order to clearly see the SFP effects on Boltzman
equations, we assume that the $\Delta L=1$ scattering processes in
Boltzman equation for heavy RH majorana neutrinos are out of
equilibrium and that the only relevant terms is the one describing
the heavy RH neutrino decays and inverse decays. Also, for the
$N_L$ Boltzman equation, we assume that all $\Delta L \neq 0$
processes induced by heavy neutrinos are out of equilibrium and
are not able to wash out any produced lepton asymmetry. Within
these approximation, the basic equations for leptogenesis
including SFP effects  are given by
\begin{eqnarray}\label{Boltzmann_mod}
\frac{dN_{N_1}}{dt}&=&-\Gamma_D N_{1} \nonumber \\
\frac{dN_{L}}{dt}&=&\epsilon \Gamma_D N_{N_1}-2 (n_F-1) \frac{d(P N_L)}{dt}\,
\end{eqnarray}
where $\Gamma_D$ represent the Direct and Inverse Decay and $N_1$ is the heavy RH Majorana neutrino density.
From \cite{Buchmuller:2004nz}, we use $\epsilon=10^{-6}$ and $\Gamma_D$ is given by
\be
\Gamma_D=\frac{1}{8 \pi} \frac{m_1 M_1}{v^2}M_1\frac{K_1(z)}{K_2(z)}.
\ee
where $K_i(z)$ are the Bessel functions, and $m_1$ is the effective light neutrino mass, $v$ is the usual electroweak symmetry breaking scale  and $M_1$ is the heavy RH neutrino mass\cite{Buchmuller:2004nz}.
The results of integrating these equations are shown in Fig \ref{fig2}. The EPT corresponds to an approximate evaluation of the Electroweak Phase transition time which corresponds to the moment when the BAU is frozen but as one can see from fig \ref{fig2}, the lepton asymmetry continue to oscillate. It is also important to notice that the total lepton asymmetry produced during leptogenesis is reduced compared to Standard scenario. This means that if we want to use leptogenesis scenario and BAU measured value to constraint neutrino masses, the effects of the magnetic fields should be taken into account as it reduced the lepton asymmetry by $50\%$. Also the uncertainties on $t_{EPT}$ implies that in presence of strong primordial magnetic fields, the uncertainties on the produced BAU could be as large as $50\%$ due to oscillating behavior of the lepton asymmetry.

\section{Conclusion}
We have studied the impact of the neutrino spin flavor precession,
induced by primordial magnetic fields, on the lepton asymmetry and
leptogenesis process. We have shown that contrary to what could be
naively expected from the weakness of the extra and intra galactic
magnetic fields at present time, primordial magnetic fields in Early Universe  could be large enough to significantly affect Leptogenesis scenario.
With such strong magnetic field at electroweak symmetry breaking time,  we have shown that the SFP
effects reduce the lepton asymmetry by around a factor $50 \%$ and
increase the uncertainties on the produced BAU as the
uncertainties on the electroweak phase transition time which
corresponds to the freezing of the BAU are important.
Even for magnetic fields too weak to modify Leptogenesis scenario, their presence induces an oscillating behavior for the lepton asymmetry at later stage in the History of the Universe, leading to lose the relation between Lepton and Baryon asymmetry as usually given in Leptogenesis models. A profile
for the magnetic fields up to time around $100$ seconds after Big
Bang is needed in order to perform more precise numerical.

\subsection*{Acknowledgements}
The work of S. K. is partially supported by the ICTP grant AC-80.
D.D. is  grateful to Conacyt (M\'exico), DAIP project (Guanajuato
University) and PIFI (Secretaria de Educacion Publica, M\'exico)
for financial support.



\begin{thebibliography}{99}
\bibliographystyle{unsrt}

\bibitem{Steigman:1976ev}
G.~Steigman.
\newblock {\em Ann.Rev.Astron.Astrophys.}, 14:339--372, 1976.

\bibitem{Steigman:2008ap}
Gary Steigman.
\newblock {\em JCAP}, 0810:001, 2008.

\bibitem{Fukugita:1986hr}
M.~Fukugita and T.~Yanagida.
\newblock {\em Phys.Lett.}, B174:45, 1986.

\bibitem{Luty:1992un}
M.A. Luty.
\newblock {\em Phys.Rev.}, D45:455--465, 1992.

\bibitem{Roulet:1997xa}
Esteban Roulet, Laura Covi, and Francesco Vissani.
\newblock {\em Phys.Lett.}, B424:101--105, 1998.

\bibitem{Buchmuller:1997yu}
W.~Buchmuller and M.~Plumacher.
\newblock {\em Phys.Lett.}, B431:354--362, 1998.

\bibitem{Buchmuller:2004nz}
W.~Buchmuller, P.~Di~Bari, and M.~Plumacher.
\newblock {\em Annals Phys.}, 315:305--351, 2005.

\bibitem{Kuzmin:1985mm}
V.A. Kuzmin, V.A. Rubakov, and M.E. Shaposhnikov.
\newblock {\em Phys.Lett.}, B155:36, 1985.

\bibitem{Matveev:1988pj}
V.A. Matveev, V.A. Rubakov, A.N. Tavkhelidze, and M.E.
Shaposhnikov.
\newblock {\em Sov.Phys.Usp.}, 31:916--939, 1988.

\bibitem{Steigman:2010zz}
Gary Steigman.
 arXiv:1008.4765 [astro-ph.CO].



\bibitem{Ahn:2012nd}
  J.~K.~Ahn {\it et al.}  [RENO Collaboration],
  arXiv:1204.0626 [hep-ex].



\bibitem{castorina}
Emanuele Castorina, Urbano Franca, Massimiliano Lattanzi, Julien
Lesgourgues,
  Gianpiero Mangano, et~al.
\newblock 2012.
\newblock 10 pages, 7 figures, 5 tables.

\bibitem{Cisneros:1970nq}
Arturo Cisneros.
\newblock {\em Astrophys.Space Sci.}, 10:87--92, 1971.

\bibitem{Okun:1986hi}
L.B. Okun, M.B. Voloshin, and M.I. Vysotsky.
\newblock {\em Sov.J.Nucl.Phys.}, 44:440, 1986.

\bibitem{Okun:1986na}
L.B. Okun, M.B. Voloshin, and M.I. Vysotsky.
\newblock {\em Sov.Phys.JETP}, 64:446--452, 1986.

\bibitem{Akhmedov:1997yv}
  E.~K.~Akhmedov,
 hep-ph/9705451.  




\bibitem{Miranda:2000bi}
O.G. Miranda, Carlos Pena-Garay, T.I. Rashba, V.B. Semikoz, and
J.W.F. Valle.
\newblock {\em Nucl.Phys.}, B595:360--380, 2001.

\bibitem{Barranco:2002te}
J.~Barranco, O.G. Miranda, T.I. Rashba, V.B. Semikoz, and J.W.F.
Valle.
\newblock {\em Phys.Rev.}, D66:093009, 2002.
\newblock new appendix added discussing the impact of the KamLAND data. This
  updates the one published in Phys.Rev.D66:093009,2002.

\bibitem{Miranda:2003yh}
O.G. Miranda, T.I. Rashba, A.I. Rez, and J.W.F. Valle.
\newblock {\em Phys.Rev.Lett.}, 93:051304, 2004.

\bibitem{Grasso:2000wj}
Dario Grasso and Hector~R. Rubinstein.
\newblock {\em Phys.Rept.}, 348:163--266, 2001.

\bibitem{Matese:1969zz}
J.J. Matese and R.F. O'Connell.
\newblock {\em Phys.Rev.}, 180:1289--1292, 1969.

\bibitem{Caprini:2001nb}
Chiara Caprini and Ruth Durrer.
\newblock {\em Phys.Rev.}, D65:023517, 2001.

\bibitem{Wang:2008vp}
Shuang Wang.
\newblock {\em Phys.Rev.}, D81:023002, 2010.

\bibitem{Kandus:2010nw}
Alejandra Kandus, Kerstin~E. Kunze, and Christos~G. Tsagas.
\newblock {\em Phys.Rept.}, 505:1--58, 2011.

\bibitem{Giovannini:1997eg}
Massimo Giovannini and M.E. Shaposhnikov.
\newblock {Primordial hypermagnetic fields and triangle anomaly}.
\newblock {\em Phys.Rev.}, D57:2186--2206, 1998.

\bibitem{Semikoz:2007ti}
V.B. Semikoz and J.W.F. Valle.
\newblock {Lepton asymmetries and the growth of cosmological seed magnetic
  fields}.
\newblock {\em JHEP}, 0803:067, 2008.

\bibitem{Semikoz:2009ye}
V.B. Semikoz, D.D. Sokoloff, and J.W.F. Valle.
\newblock {\em Phys.Rev.}, D80:083510, 2009.

\bibitem{Schechter:1981hw}
J.~Schechter and J.W.F. Valle.
\newblock {Majorana Neutrinos and Magnetic Fields}.
\newblock {\em Phys.Rev.}, D24:1883--1889, 1981.

\bibitem{Lim:1987tk}
Chong-Sa Lim and William~J. Marciano.
\newblock {\em Phys.Rev.}, D37:1368--1373, 1988.

\bibitem{Ichiki:2011ah}
Kiyotomo Ichiki, Keitaro Takahashi, and Naoshi Sugiyama.
\newblock {\em Phys.Rev.}, D85:043009, 2012.

\bibliographystyle{unsrt}
\end{thebibliography}
\end{document}